\documentclass{DISproc}

\begin{document}
\title{The High Energy Jets Framework}

\author{{\slshape Jennifer M.~Smillie}\\[1ex]
University of Edinburgh, Mayfield Road, Edinburgh, EH9 3JZ, UK}

\contribID{http://indico.cern.ch/contributionDisplay.py?contribId=233\&confId=153252}


\maketitle

\begin{abstract}
  High Energy Jets provides an all-order description of wide-angle QCD
  emissions, resumming the leading-logarithmic contributions in the high-energy
  limit.  In this contribution, we briefly summarise the approach and its
  implementation in a flexible Monte Carlo event generator.  We discuss
  comparisons between HEJ and recent LHC data and then go on to probe the
  similarities and differences in the results obtained from High Energy Jets and
  other theoretical frameworks in inclusive dijet and W+dijet production.
\end{abstract}

\section{Introduction to High Energy Jets (HEJ)}

Accurate theoretical descriptions of multi-jet production are of key importance
to physics at the Large Hadron Collider.  This is our first opportunity to test
our theoretical understanding of QCD at these high energies, and this will be
key to reaching the full potential of the LHC physics programme.  For example,
the results of comprehensive analyses of multi-jet radiation with current data
will be used when applying a jet veto to the production of a Higgs boson in
association with jets.

It has already been seen in the \unit{7}{\TeV} LHC data that the ratio between
inclusive $(n+1)$-jet rates and inclusive $n$-jet rates can be large.  While
this is true for the ratios of the cross sections, the effect is particularly
large in certain key regions of phase space including high momentum regions (see
e.g.~high $H_T$ in \cite{Aad:2012en}) or events where there is a large rapidity
separation between jets (see e.g.~\cite{Aad:2011jz}).

Motivated by the large impact of higher order corrections, the High Energy Jets
(HEJ) framework~\cite{Andersen:2009nu,Andersen:2009he,Andersen:2011hs} provides
an all-order resummation of the dominant (leading-log) contributions to
wide-angle, hard QCD radiation in the High Energy limit.  In this limit,
scattering amplitudes factorise into rapidity-ordered pieces.  This structure
allows an extremely efficient description of many-particle hard-scattering
matrix elements.  This forms the basis of the HEJ description which has been
developed for the production of jets, and also $W$, $Z$ and Higgs boson
production in association with jets.  The High Energy limit can be stated as
\begin{equation}
  \label{eq:mrk}
  s_{ij}\to \infty\ \forall\ \{i,j\}, \quad |p_{\perp, i}|\ {\rm fixed},
\end{equation}
where $i,j$ label outgoing quarks and gluons.  In practice this corresponds to
wide-angle QCD emissions and may be stated equivalently in terms of pairwise
rapidity differences becoming large while transverse momenta components remain
finite.  This is in contrast to the soft and collinear emissions which are
included in a parton shower resummation.  A complete jet description can be achieved by
consistently merging the two approaches~\cite{Andersen:2011zd}.

The derivation of the building blocks of the HEJ framework has been described
in detail in~\cite{Andersen:2009nu,Andersen:2009he}.  The implementation of these
for multi-jet production in a fully flexible event generator is described
further in~\cite{Andersen:2011hs}, and the generator itself is publicly
available at \texttt{www.cern.ch/hej}.  

Predictions from HEJ have been used in analyses by ATLAS~\cite{Aad:2011jz} and
by CMS~\cite{Collaboration:2012gw,Chatrchyan:2012pb}.  The ATLAS study was a
study of jet radiation with a jet veto across a wide range of transverse momenta
and rapidities.  HEJ gave a consistently good description of data throughout.
Discrepancies were only seen in cases where cuts had induced a large hierarchy
of transverse momentum scales, as this evolution is not systematically included
in the parton-level predictions.  In the central-forward CMS
study~\cite{Collaboration:2012gw} which separated the jets in rapidity, HEJ
again gave a good description of data where more traditional approaches
performed less well.  In a subsequent jet study~\cite{Chatrchyan:2012pb}, the
HEJ predictions showed slight deviation from data at large rapidity differences;
work is ongoing to evaluate the uncertainties in this case.  

Overall, HEJ has given an excellent description of early data, and in some cases
has out-performed other more standard approaches.  This underlines the
importance of the higher order contributions included in HEJ.  In the rest of
this contribution, we probe to what extent data could probe the differences
between the HEJ approach and that of other theoretical frameworks which are
built upon fixed-order matrix elements.

\section{Comparisons Between Theoretical Approaches}

We begin by comparing HEJ and POWHEG~\cite{Alioli:2010xd,Alioli:2010xa}
predictions for dijet production.  The HEJ framework is an all-order resummation
of wide-angle QCD radiation which includes, for events which result in four or
fewer jets, matching to leading-order matrix elements.  In contrast, the POWHEG
description of multi-jet production begins with a next-to-leading order (NLO)
matrix element, which is then supplemented with a resummation from a parton
shower.  It is surprising, then, that the predictions from the two approaches
have been seen to be very similar (see \cite{Aad:2011jz}).  The extent to
which these descriptions can be distinguished was studied recently
in~\cite{Alioli:2012tp}.

In order to implement cuts which do not induce a large hierarchy in
transverse momentum between the jets, a minimal set is used:
\begin{equation}
  \label{eq:cuts}
  p_{\perp,j}>35\ {\rm GeV},\quad p_{\perp,j_1}>45\ {\rm GeV},\quad |y_j|<4.7.
\end{equation}
The additional cut on the hardest jet is in order to allow a meaningful
comparison with the pure NLO calculation.  Neither the POWHEG or HEJ
descriptions suffer from an instability in the presence of symmetric cuts.

\begin{figure}[tbp]
  \centering
  \includegraphics[width=0.48\textwidth]{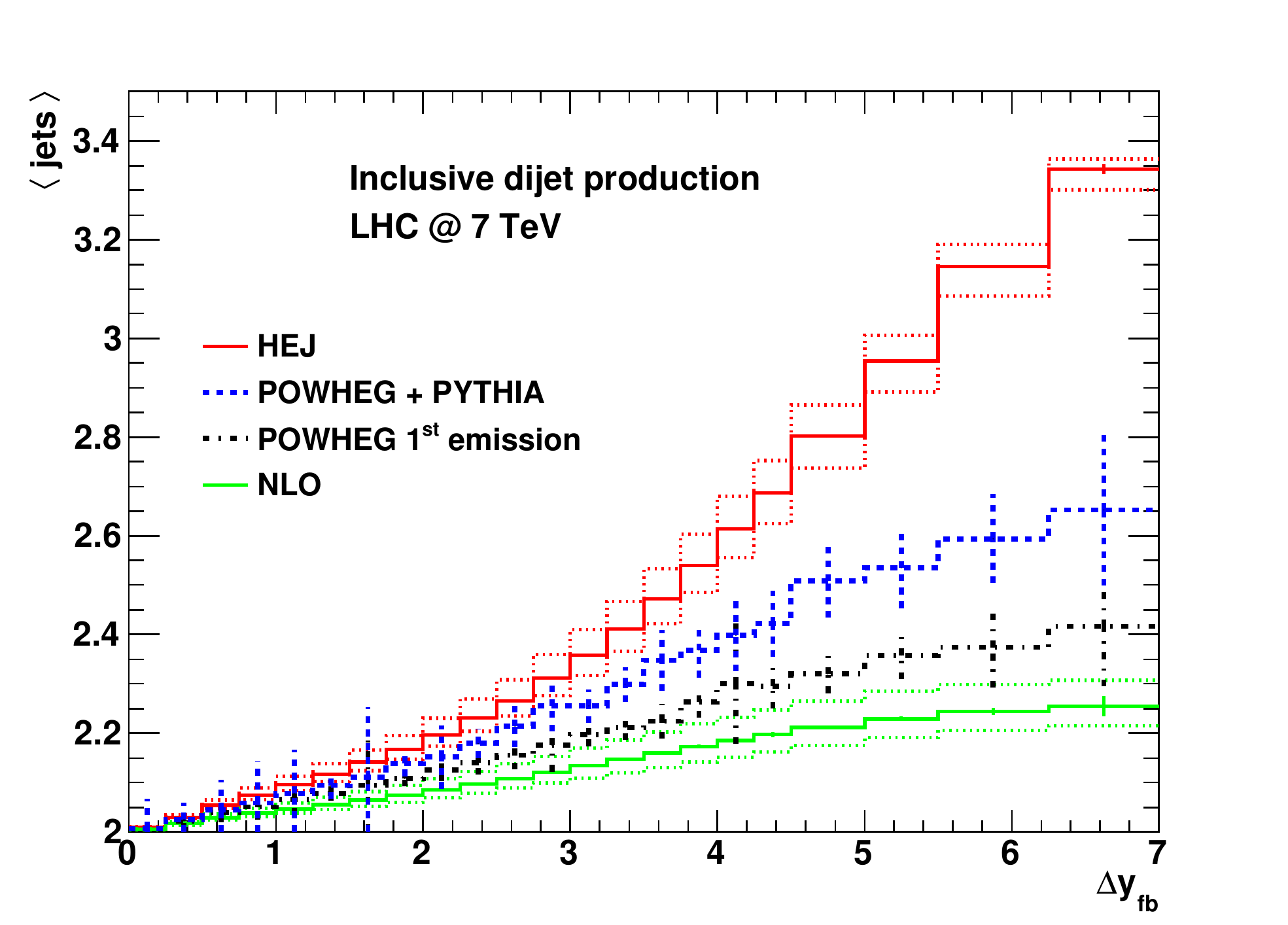}
  \includegraphics[width=0.48\textwidth]{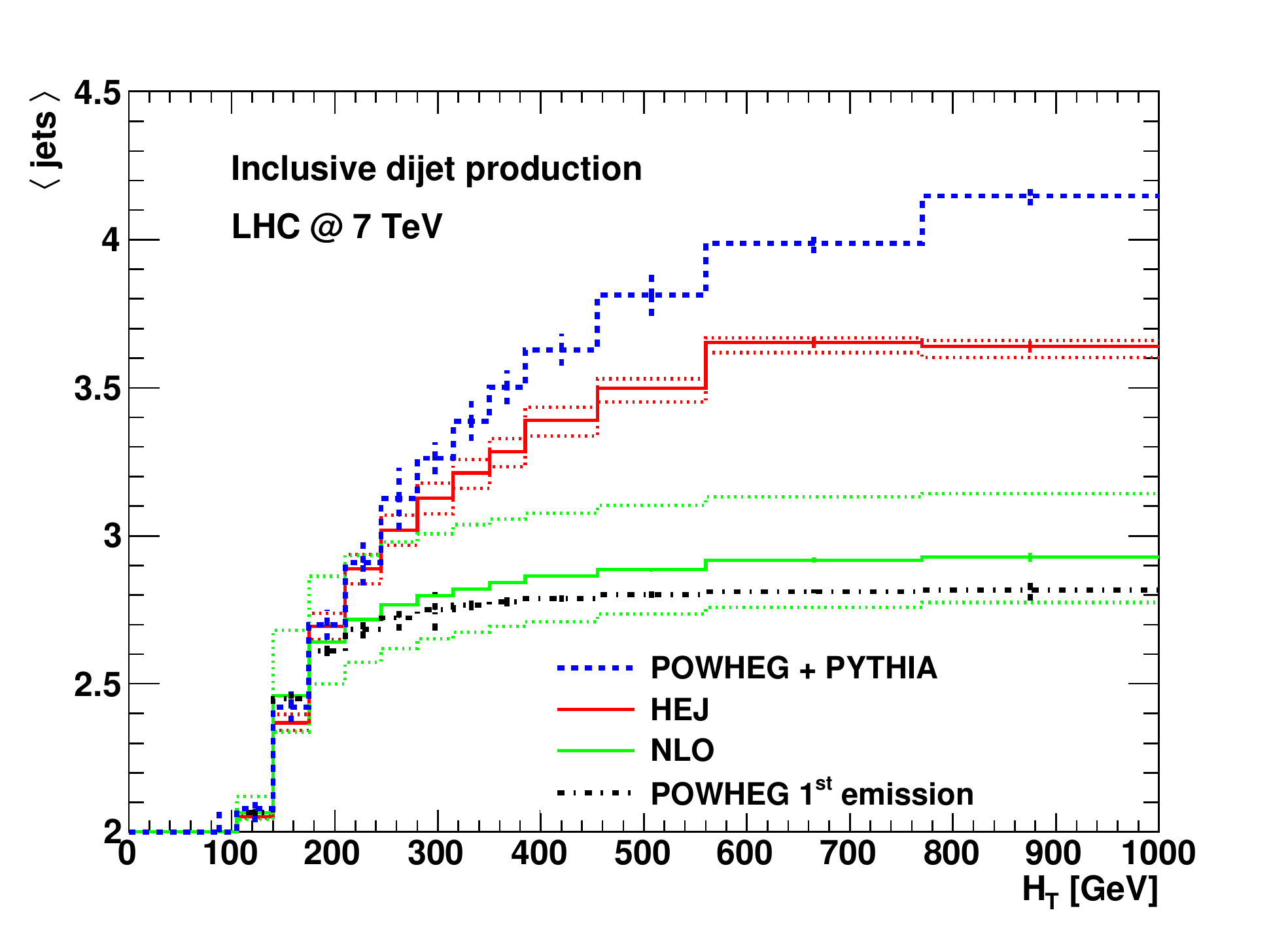}
  \caption{The average number of jets as a function of (left) the difference in
    rapidity between most forward and most backward jet, and (right) as a
    function of $H_T$.  Plots taken from~\cite{Alioli:2012tp}.}
  \label{fig:avgjets}
\end{figure}

The left plot in figure~\ref{fig:avgjets} shows the predictions for the average
number of jets as a function of the rapidity difference between the most forward
and most backward jet in each event, $\Delta y_{fb}$.  The bands around the HEJ
and NLO predictions indicate the result of varying the renormalisation and
factorisation scales by a factor of two in each direction.  The vertical lines
indicate statistical uncertainty.  In this plot, all the predictions show an
increase in the average number of jets with $\Delta y_{fb}$, with the largest
increase being seen in the HEJ prediction, as expected.  The lowest prediction
comes from the pure NLO calculation, followed by the POWHEG first emission, then
the full POWHEG+PYTHIA shower, which increases to a value around 2.6 for $\Delta
y_{fb}=7$.

The same variable is shown in the right plot in figure~\ref{fig:avgjets}, but
now as a function of $H_T$.  It is immediately clear from the different
behaviour that a different region of phase space is now being probed.  As $H_T$
increases, the largest prediction now comes from the POWHEG+PYTHIA prediction
which peaks above an \emph{average} value of 4 jets per event, which is
remarkably high for an inclusive dijet sample.  The HEJ prediction levels off a
little below this around 3.6. The NLO and POWHEG 1st emission predictions are
restricted to lie below 3, and both reach values close to that.

The differences in the predictions here appear to be significant enough that one
could hope to distinguish between the approaches with LHC data.  Other variables
were also studied in~\cite{Alioli:2012tp}, which showed smaller differences.
For example, when a measure of the azimuthal decorrelation of the jets
(which results from hard radiation) is studied as a function of $\Delta y_{fb}$,
the predictions from HEJ and POWHEG+PYTHIA are extremely similar until values of
$\Delta y_{fb}>6$.

\begin{figure}[btp]
  \centering
  \includegraphics[width=0.48\textwidth]{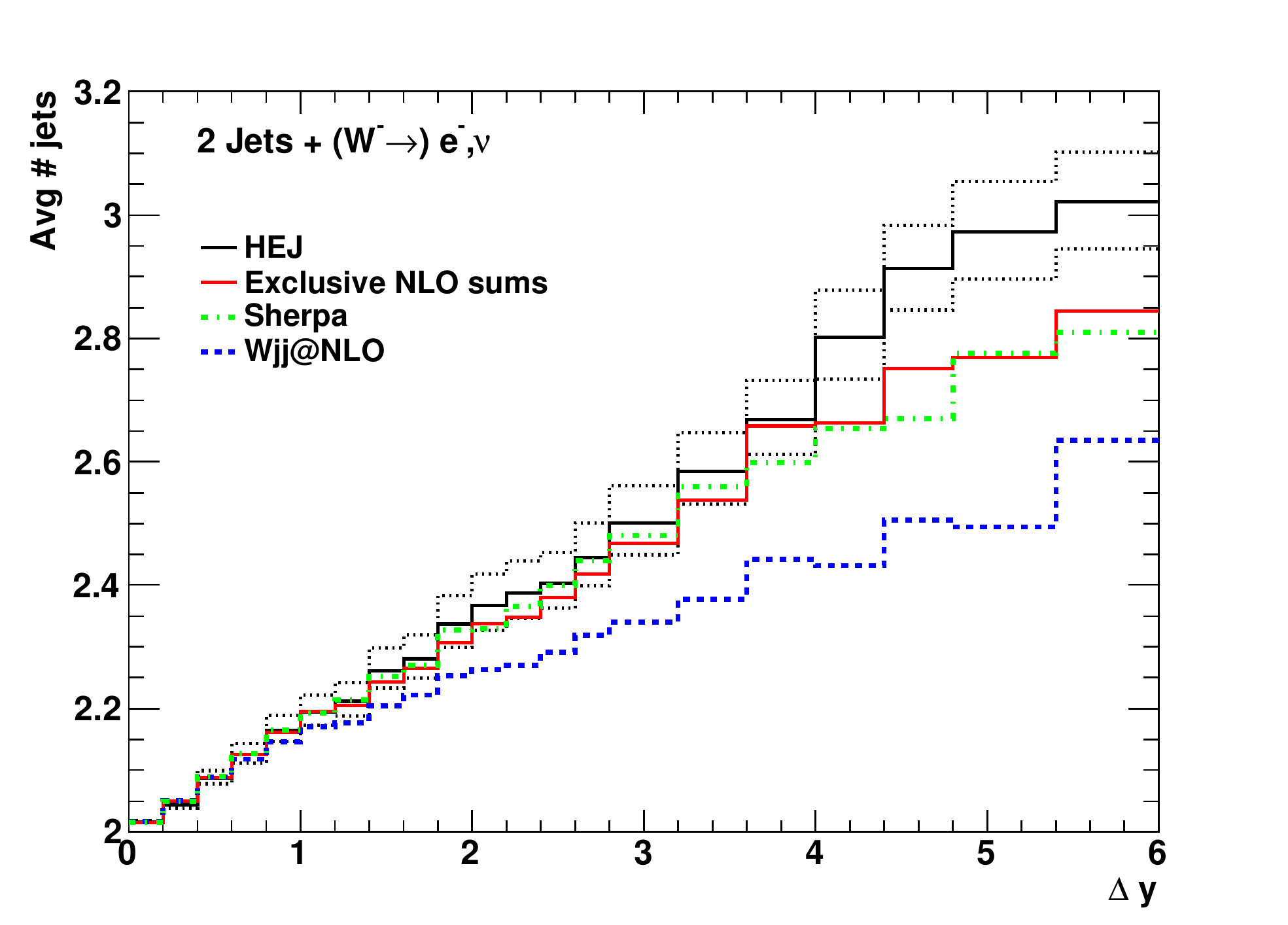}
  \includegraphics[width=0.48\textwidth]{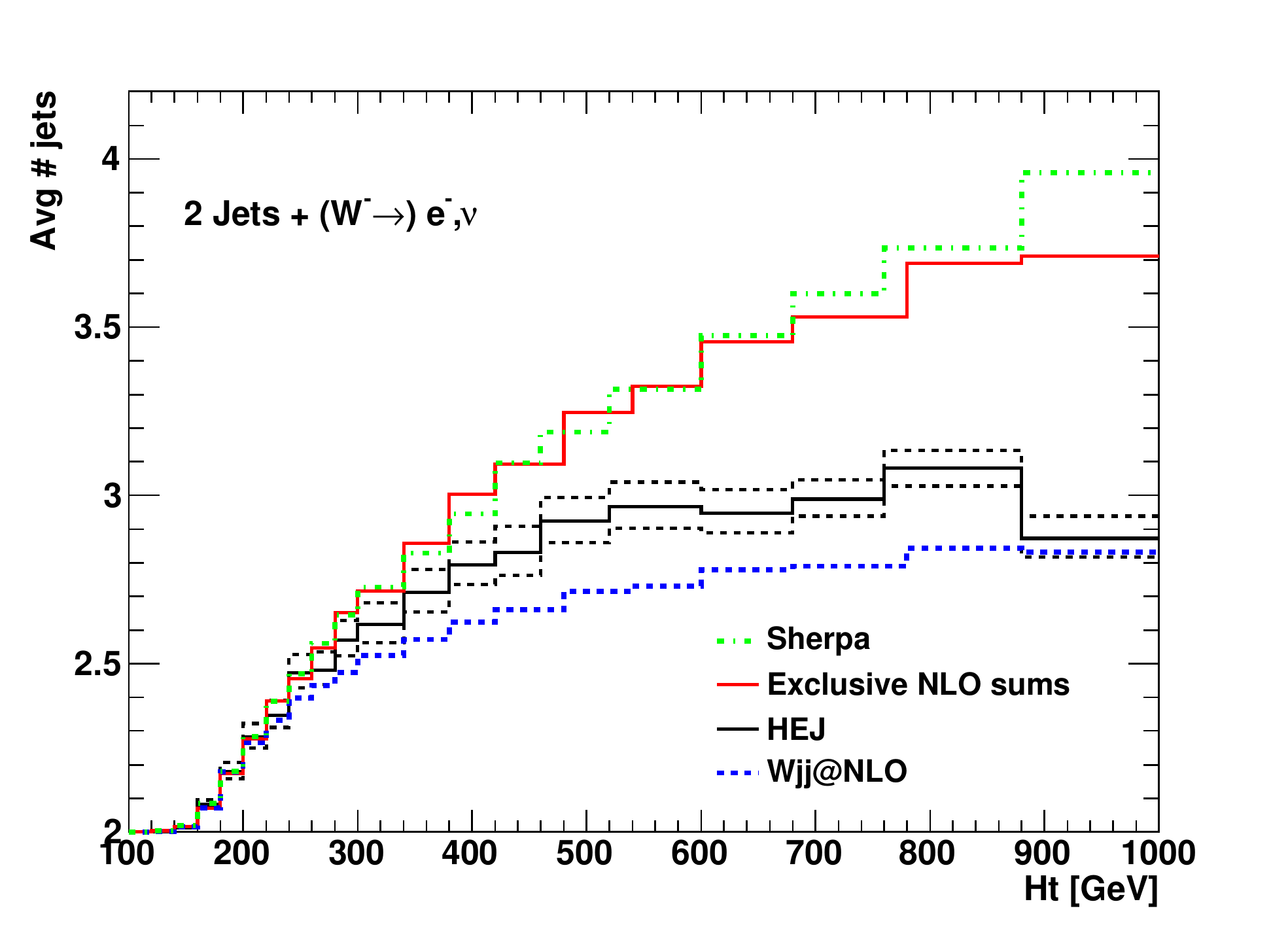}
  \caption{A comparison of the predictions for the average number of jets in an
    inclusive $W$+dijet sample from different theoretical descriptions as a
    function of (left) the difference in rapidity between most forward and most
    backward jet, and (right) as a function of $H_T$. Plots taken
    from~\cite{Maestre:2012vp}.}
  \label{fig:Wavgjets}
\end{figure}

A related study has been performed in the context of $W$ boson production in
association with jets in~\cite{Maestre:2012vp}.  Here, predictions from four
theoretical approaches were compared: NLO and a merged NLO sample both from
BlackHat~\cite{Berger:2009zg,Berger:2009ep,Berger:2010zx}, a merged
matrix-element plus parton shower sample from
Sherpa~\cite{Hoeche:2009rj,Hoeche:2009xc,Carli:2010cg} and
HEJ~\cite{Andersen:2012gk}.  Figure~\ref{fig:Wavgjets} shows the predictions for
the average number of jets now for this process.  In the left plot, it can again
be clearly seen that the predictions all rise with $\Delta y_{fb}$.  In this
case, there is a high level of agreement between the predictions until large
values of $\Delta y$, with only the pure NLO prediction lying slightly below.  

In the right-hand plot of figure~\ref{fig:Wavgjets}, the average number of jets
for $W$+jets is shown as a function now of $H_T$.  Here, the SHERPA and
Exclusive NLO sums predictions give the highest value for large $H_T$, peaking
around a value of 4.  The HEJ prediction here is lower, and closer to the pure
NLO result around 3.  It should be possible to distinguish between these with
data.

\section{Summary}
The High Energy Jets framework provides an alternative method to describing
multi-jet production, which is based on an all-order resummation of hard,
wide-angle QCD radiation.  It has already been seen to give a good description
of early LHC data.  Analyses which may be able to distinguish between different
theoretical descriptions in jets and W+jets have been discussed.  For example,
the average number of jets as a function of $H_T$ shows large differences
between different theoretical approaches for both processes.

\section*{Acknowledgements}

JMS thanks the conference organisers and the conveners of the \emph{Hadronic
  Final States} Working Group for an interesting and fruitful week.  JMS is
funded by the UK Science and Technology Facilities Council (STFC).

{\raggedright
\begin{footnotesize}
  \bibliographystyle{DISproc}
  \bibliography{smillie_jennifer.bib}
\end{footnotesize}
}


\end{document}